# Pentacene islands grown on ultra-thin SiO$_2$

*B. R. Conrad, W. G. Cullen, B. C. Riddick, E. D. Williams*

Physics Department and Materials Research Science and Engineering Center,

University of Maryland

College Park, MD 20742-4111, USA

Corresponding author: edw@umd.edu

Abstract

Ultra-thin oxide (UTO) films were grown on Si(111) in ultrahigh vacuum at room temperature and characterized by scanning tunneling microscopy. The ultra-thin oxide films were then used as substrates for room temperature growth of pentacene. The apparent height of the first layer is 1.57±0.05 nm, indicating "standing up" pentacene grains in the thin-film phase were formed. Pentacene is molecularly resolved in the second and subsequent molecular layers. The measured in-plane unit cell for the pentacene (001) plane (**ab** plane) is **a**=0.76±0.01 nm, **b**=0.59±0.01 nm, and γ=87.5±0.4°. The films are unperturbed by the UTO's short-range spatial variation in tunneling probability, and reduce its corresponding effective roughness and correlation exponent with increasing thickness. The pentacene surface morphology follows that of the UTO substrate, preserving step structure, the long range surface rms roughness of ~0.1 nm, and the structural correlation exponent of ~1.

Recent growth in the field of organic semiconductors is indicative of the continued interest in their unique physical and chemical properties [1, 2]. However, limited understanding of electronic transport in these systems, in particular the poorly understood effects of conduction channel morphology at the molecular scale, hinders application development [3-5]. Due to its robust ordering on a variety of substrates, relatively high mobility, and simple chemical structure, pentacene (Pn) has become the effective benchmark for organic thin film transistors [6] and other applications [7]. For this model organic semiconductor, it is well known that substrate topography, roughness, and trapped charges highly influence growth and morphology, as well as device characteristics [8-10]. These effects are well documented for the standard substrate used in electronic devices and transport studies, which is $SiO_2$. Ideally, molecular resolution imaging techniques such as STM would be used to probe the Pn interactions with the $SiO_2$ substrate, as well as the effects on the Pn crystal structure and morphology. However, tunneling measurements are precluded on dielectric substrates [11]. Here we demonstrate that this problem can be addressed by using ultra-thin layers of $SiO_2$ on Si as model substrates.

The crystal structure of Pn on an $SiO_2$ substrate [12-14] is well known. Pn films have also been imaged with molecular resolution on metals [15, 16] and the ultra-thin insulating layers NaCl/Cu(111) [17] and Bi/Si(111) [18, 19]; however, the Pn crystal structure on an insulating oxide surface has not been imaged. To accomplish imaging, we use the approach of current commercial CMOS devices which utilize an ultra-thin oxide (UTO) approximately as thick as native oxides on the growth substrate [20, 21]. The surfaces of such silicon oxide layers are known to be smooth in comparison with thick $SiO_2$, because the thin $SiO_2$ closely follows the morphology of the atomically clean Si substrate precursor [22, 23]. Thus the UTO substrate also allows us to probe the effects of the relative roughness of thin and thick [24] SiO$_2$ substrate layers on the growth and crystallinity of the Pn films.

The experiments were conducted in an ultrahigh vacuum (UHV) chamber (base pressure $\sim 4\times10^{-11}$ Torr) with a variable temperature Omicron scanning tunneling microscope (STM). The n-doped silicon wafers (<0.1 Ohm-cm) were misoriented by 0.5º toward the [$2\bar{1}\bar{1}$] direction. The Si surface was prepared by several 5s flashes at 1530 K with subsequent cooling at a slow rate (~ 30ºC/min) though the (1x1)-to-(7x7) phase transition. The Si substrate was heated resistively with direct current while the temperature was measured via an infrared pyrometer. The ultra-thin oxide layers were formed by exposing the atomically clean Si(111)-(7x7) to $2.4\times10^6$ L of O$_2$ at room-temperature. The samples were then outgassed at 300ºC in UHV and imaged afterward to confirm oxide quality. The Pn films were grown in an attached chamber at a base pressure $<1\times10^{-9}$ Torr at a rate of 0.5 ML/min, with flux measured by a water-cooled quartz microbalance (Leybold Inficon). All STM measurements presented were performed in constant current mode (<40pA) with electrochemically etched tungsten tips and a sample bias of 2.7 to 3.0V.

Figure 1a shows a representative (500$nm$)$^2$ STM image of the UTO layer. The linear features in the image are single or double atomic silicon steps whose heights, 0.31 nm and 0.63 nm respectively, agree with STM measurements of the clean Si(111). No Si(111)-(7x7) structure can be identified in STM images and low energy electron diffraction (LEED) measurements show no 7x7 reconstruction remaining after exposures larger than $1\times10^6$ L of O$_2$. This confirms complete oxidation of the surface, and previous work suggests a film up to 1nm thick [25, 26]. Figure 1b shows a representative (100$nm$)$^2$ STM image of the UTO surface. The observed image nonuniformity of the silicon terraces has been reported elsewhere [27, 28], and has been attributed to variations in the interfacial nucleation of the oxide and corresponding variability in the tunnel conductance, rather than variations in surface height [29, 30]. Our observation that the image nonuniformity is independent of tunneling conditions supports this conclusion. When a single layer of Pn is grown on the UTO as shown in the $(100 nm)^2$ STM image of Figure 1c, the image nonuniformities are similar to those seen on the UTO terraces in Figure 1b.

The initial stage of Pn growth on a UTO film is illustrated in Figure 2a, which is an *in situ* $(500 nm)^2$ STM image of a thermally grown pentacene island. The height of the first layer pentacene islands is measured to be 1.57±0.05 nm, confirming a "standing-up" configuration mostly normal to the surface, in agreement with Pn thin film phase measurements [8, 9, 31], as shown in Table 1. As Figure 2a illustrates, the Pn islands exhibit dendritic growth characteristics, which are similar to growth on thicker and rougher $SiO_2$ substrates [6, 10, 24]. The underlying structure of the stepped Si surface can be seen in the figure as the diagonal linear features and we observe Pn growth to be independent of the steps despite their 0.31 nm height. Figure 2b shows a $(500 nm)^2$ STM image of an incomplete second layer of Pn. Similar growth structures are observed for second and subsequent Pn layers with increasing three-dimensional growth due to an Ehrlich-Schwoebel-type barrier [6, 10] common in many organic systems [32]. The height of the second and third Pn layers is measured with STM to be 1.58±0.05 nm, equal to that of the first layer as shown in Table 1

Figure 3a shows a higher resolution STM image $(20 nm)^2$ of the first layer of Pn showing nonuniformities similar to those seen in Figure 1c. Molecular resolution cannot be achieved in the Pn layer in immediate contact with the UTO, independent of tunneling or scanning parameters. The cause could be a disordered first Pn layer or some unidentified electronic coupling between Si/SiO$_2$ and Pn. However, Figure 3b shows a higher resolution STM image $(20nm)^2$ of the second layer of pentacene. A periodic molecular-scale structure can be seen, overlaid with a longer scale light/dark variation similar to the nonuniformity seen in Figures 1b, 1c and 3a. The lattice parameters of the periodic structure were measured from the molecular resolution images such as the $(20nm)^2$ STM image seen in Figure 3b. The in-plane unit cell values obtained from several STM measurements are **a**=0.76±0.01 nm, **b**=0.59±0.01 nm, and γ=87.5±0.4°. These measurements confirm that the second layer structure is the Pn(001) plane (**ab** plane) in the thin film phase, or Pn polymorph IV [9], as is commonly observed in Pn films on thick SiO$_2$ systems [4, 13, 14].

The morphology observed by STM can be attributed to variations conductance variations for length scales smaller than a single terrace, and to the density of Si steps for length scales larger than a single terrace. To compare the short and long-range roughness characteristics [33-35], the 2D STM height-height correlation function, $g(r) = \langle (z(r_o + r) - z(r_o))^2 \rangle$, was determined using the STM height measurements z(r) for the atomically clean and Pn-covered UTO surfaces. The correlation functions are observed to behave as $g(r) \sim r^{2H}$ with two separate signatures at large and small length scales. For length scales larger than L = 5 nm, e.g. for image sizes larger than the average step-step separation, we observe a correlation exponent 2H~1 with a correlation length of ξ=22 ± 2 nm, similar to the long range morphology of Si(111) and thick SiO$_2$ films [34-37]. Another measure of the surface roughness is the root mean square (RMS) roughness. For images larger than 300nm, which include many terraces, the RMS roughness is 0.109±0.014 nm for the UTO layer and 0.099±0.007 nm for the second layer of Pn, compared with a typical

RMS roughness of 0.3 nm for thick $SiO_2$. This indicates that on the UTO, the long range roughness is dominated by the Si step density and the over-layers do not develop independent roughness beyond that of the growth substrate.

In contrast to the large scale analysis, Figure 4 shows short-scale (analysis area 40 x 40 $nm^2$) representative radially averaged 2D STM height-height correlation functions for the UTO terraces, continuous Pn layers on the UTO terraces, and two continuous Pn layers on UTO terraces. At these short length scales the measured 2D STM height-height correlation exponent for individual terraces, as shown in Table 1, is larger than at long length scales because measurements are dominated by the apparent height differences due to the short length scale conductance variations on the terraces. The correlation length $\xi = 5.0 \pm 0.5$ nm, for small length scales does not change with Pn thickness, as is expected since the length scale of the nonuniformities does not change with Pn thickness. The short-range STM correlation exponent, $2H=1.60\pm0.03$, is much larger than the long range exponent ($2H\sim1$), or of the short-range exponent measured using AFM [38]. The apparent decrease of the correlation exponent from $2H=1.60\pm0.03$ for the UTO to $2H=1.38\pm0.03$ for the second layer Pn grain suggests that the Pn over-layers are partially smoothing the conductance variations due to the oxide. The measured short scale, or terrace, RMS roughnesses in Table 1 also show a continuous decrease in apparent terrace roughness from $0.076\pm0.011$ nm for the UTO to $0.050\pm0.002$ nm for the third Pn layer despite the observed molecular order, of amplitude ~20 pm, of the second and subsequent Pn films. The conformal growth of the Pn and its ability to shield conductance variations suggests its usefulness as an interlayer on an oxide gate dielectric or electrodes [7].

In conclusion, we have imaged and characterized pentacene grown on an ultra-thin silicon oxide film with STM. Pentacene's growth on UTO is qualitatively similar to that on thicker oxides, which indicates that UTO is a useful substrate for studying pentacene as well as other organic semiconductors under conditions relevant to devices. Molecular resolution STM imaging is achieved for a pentacene film on $SiO_2$ for the first time and the thin film phase crystal structure of the pentacene grains appears at the second layer of growth. The STM-image roughness of the ultra-thin $SiO_2$ layer is shown to have distinctly different character for long and short length scales. The long scale STM surface roughness is about three times smaller than that of thick oxides, and does not increase with growth of pentacene over-layers, which is conformal. On individual terraces, *e.g.* at short length scales, the apparent surface roughness diminishes as a function of the number of Pn over layers indicating damping of the conduction variation, in contrast to the conformal growth over substrate steps.

Acknowledgements

This work was supported by NIST under contract no. 70NANB6H6138 with research infrastructure supported by the UMD-CNAM and NanoCenter. The NSF-MRSEC SEF was used in obtaining the data presented.

Table 1 The measured Pn layer height, Pn lattice constants a, b and γ, short scale RMS roughness (L < 30 nm), long scale RMS roughness(L > 300 nm), and 2D correlation exponent 2H for the oxide layer (oxide) as well as the first, second, and third Pn layers (Pn 1, Pn 2, Pn 3).

|  | Layer height (nm) | a (nm) | b (nm) | γ (°) | Short range RMS roughness (nm) | Short range STM correlation exponent 2H | Long range RMS roughness * (nm) |
|---|---|---|---|---|---|---|---|
| Oxide | - | - | - | - | 0.076±0.011 | 1.60±0.03 | 0.109±0.014 |
| Pn 1 | 1.57±0.05 | - | - | - | 0.060±0.001 | 1.50±0.03 | - |
| Pn 2 | 1.58±0.05 | 0.76±0.01 | 0.59±0.01 | 87.5±0.4 | 0.050±0.001 | 1.38±0.03 | 0.099±0.008 |
| Pn 3 | 1.58±0.05 | 0.76±0.01 | 0.59±0.01 | 87.5±0.4 | 0.050±0.002 | - | - |

* The long range height roughness exponent is 2H~1.

Figures

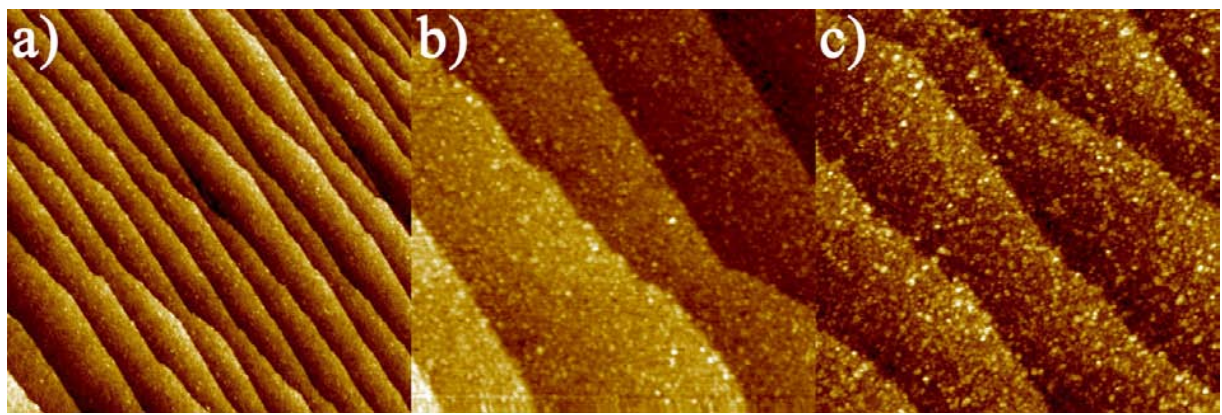

**Figure 1:** (color online) a) A $(500nm)^2$ STM image of ultra thin film oxide on a stepped Si(111) surface with a pixel size of 1.25nm measured at room temperature. Single step heights are 0.31 nm. b) A $(100nm)^2$ STM image of ultra thin film oxide on a stepped Si(111) surface with a pixel size of 0.39 nm measured at room temperature. Single step heights are 0.31 nm. c) A $(100nm)^2$ STM image of a continuous pentacene film on a similar ultra thin oxide film.

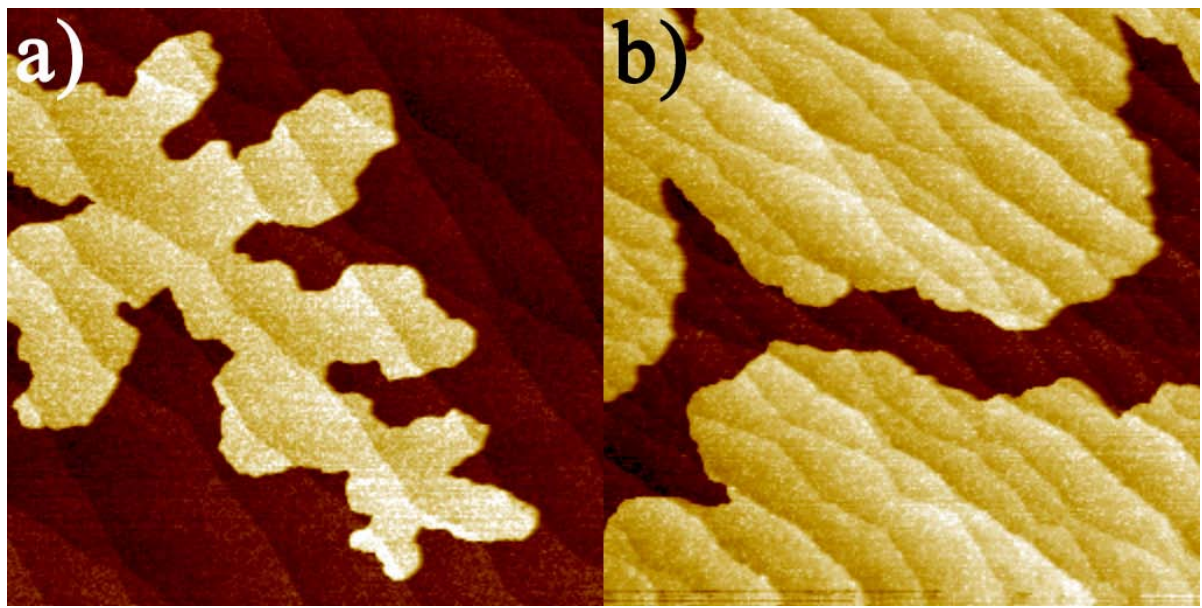

**Figure 2:** (color online) a) A $(500nm)^2$ STM image of a pentacene island on an ultra thin film oxide on a stepped Si(111) surface with a pixel size of 1.95 nm measured at room temperature. Single step heights are 0.31 nm on both the oxide and the pentacene island. b) A $(500nm)^2$ STM image of a pentacene grains on a complete pentacene monolayer on a similar ultra thin film oxide with a pixel size of 1.95 nm measured at room temperature. All single step heights are measured to be 0.31 nm.

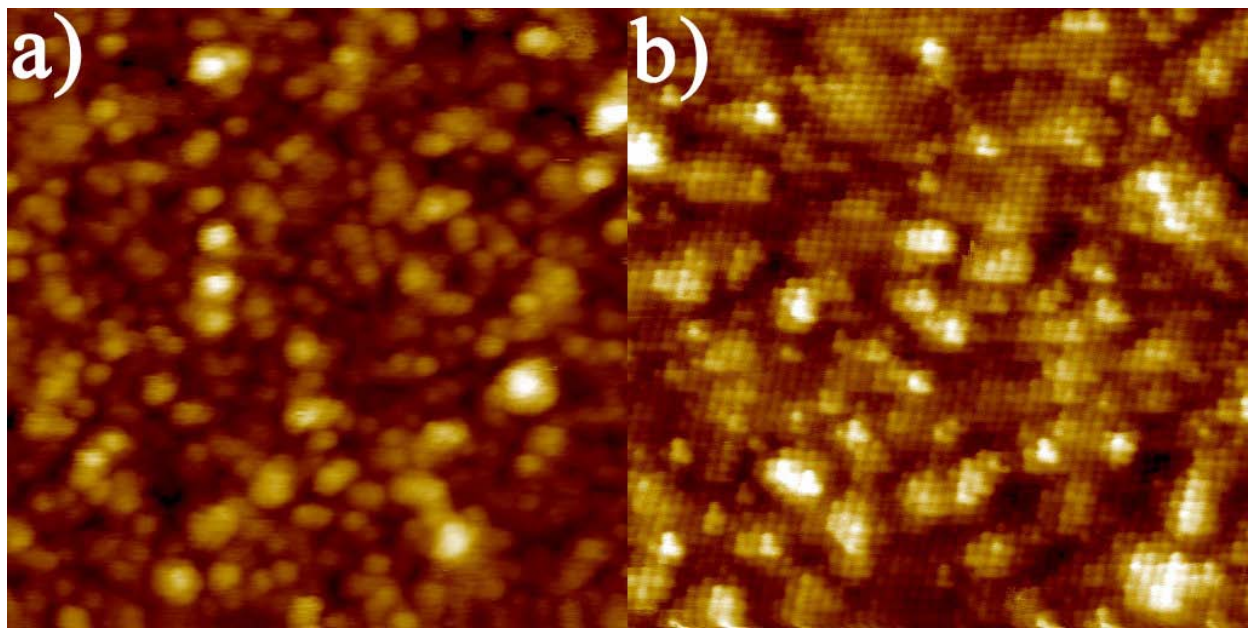

**Figure 3:** (color online) a) A $(20nm)^2$ STM image of a pentacene island on an ultra thin film oxide terrace with a pixel size of 0.04 nm and a height range of 530 pm measured at room temperature. b) A $(20nm)^2$ STM image of a pentacene island on complete pentacene monolayer on a similar ultra thin film oxide terrace with a pixel size of 0.04 nm and a height range of 380 pm measured at room temperature. The peak-to-peak modulation amplitude due to the periodic structure of the pentacene is ~20 pm.

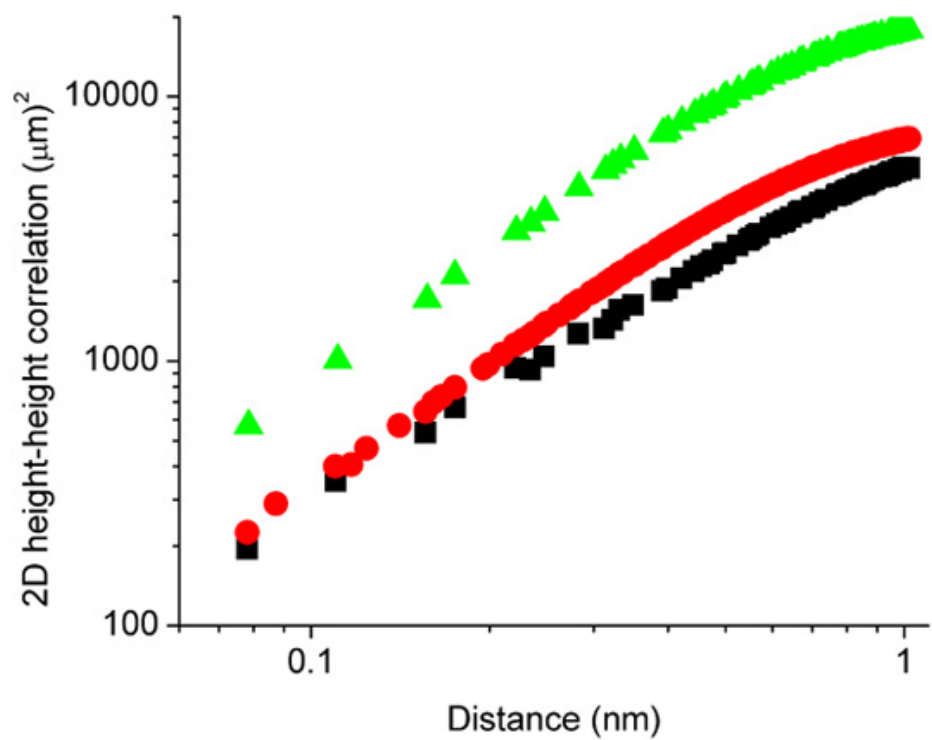

Figure 4: Short range 2D STM correlation functions of a representative UTO, a singly pentacene-covered UTO, and a doubly pentacene-covered UTO. Fits of correlation functions to the functional form $g(r) \sim r^{2H}$ yield short-range STM correlation exponents 2H=1.60±0.03 for the UTO (green triangles), 2H=1.50±0.03 for the first pentacene layer (red circles), and 2H=1.38±0.03 for the second pentacene layer (black squares).